\documentclass[english]{sbrt}
\usepackage{amsmath}
\usepackage{amssymb}
\usepackage[english]{babel}
\usepackage{svg}
\usepackage[none]{hyphenat}
\usepackage{float}
\usepackage{cite}
\PassOptionsToPackage{hyphens}{url}
\usepackage{url}
\usepackage{breakurl}

\begin{document}

\bstctlcite{IEEEexample:BSTcontrol}
\title{Comparative Performance Analysis of OTFS and OFDM Modulations for Mobile Wireless Communications}

\author{J. Marcos Leal B. Filho, Danilo de S. Pena, Marília C. Muniz, Álvaro A. M. de Medeiros, Vicente A. de Sousa Jr.
\thanks{J. Marcos Leal B. Filho, Marília C. Muniz e Vicente A. de Sousa Jr are from Federal University of Rio Grande do Norte, E-mails: jose.filho.048@ufrn.edu.br, marilia.muniz.124@ufrn.edu.br, vicente.sousa@ufrn.br. Danilo de S. Pena is from the Faculty of Engineering, University of Porto, E-mail: danilopena.ba@gmail.com. Álvaro A. M. de Medeiros is from Federal University of Juiz de fora, E-mail: alvaro.medeiros@ufjf.br. This study was funded in part by Coordenação de Aperfeiçoamento de Pessoal de Nível Superior - Brasil (CAPES) - Financing Code 001. The proof of concept simulations provided by this paper was supported by High-Performance Computing Center at UFRN (NPAD/UFRN).}}

\maketitle

\markboth{XLIII BRAZILIAN SYMPOSIUM ON TELECOMMUNICATIONS AND SIGNAL PROCESSING - SBrT 2025, SEPTEMBER 29TH TO OCTOBER 2ND, NATAL, RN}{}

\begin{abstract}

This paper provides a quantitative performance comparison between Orthogonal Time Frequency Space (OTFS) and Orthogonal Frequency-Division Multiplexing (OFDM) modulation schemes, focusing on mobile wireless communication scenarios. We evaluate and compare both schemes based on critical communication scenarios and configurations such as mobility levels, modulation orders, multipath environments, and equalizers. The study systematically identifies conditions where OTFS and OFDM each exhibit optimal performance. Results from simulations demonstrate that OTFS outperforms OFDM consistently for high mobility scenarios and multipath environments. Depending on the modulation order, the performance gap between OTFS and OFDM might be very close or many orders of magnitude. Moreover, at low and mid values of SNR, the non-linear equalizer performs better than traditional linear equalizers for OTFS.

\end{abstract}

\section{Introduction}

Wireless communication systems face significant challenges in environments characterized by high mobility, where traditional modulation techniques often struggle to maintain robust and reliable performance. Orthogonal Frequency-Division Multiplexing (OFDM) has long served as a fundamental modulation technology for numerous wireless standards due to its efficient spectrum usage and robustness against multipath fading. However, OFDM transmits symbols in the time-frequency domain, assigning each subcarrier per OFDM symbol, which makes it highly susceptible to performance degradation in rapidly time-varying channels. High Doppler shifts cause Inter-Carrier Interference (ICI), significantly affecting OFDM's performance in high-mobility scenarios~\cite{rappaport2020wireless}.

On the other hand, Orthogonal Time Frequency Space (OTFS) modulation maps the information symbols in a two-dimensional delay-Doppler grid, effectively converting the time-varying wireless channel into an approximately invariant two-dimensional response~\cite{hadani2017otfs}. This representation allows OTFS symbols to benefit from the full time-frequency diversity of the channel, thus offering robustness against delay and Doppler spread~\cite{hong2022delay}. In essence, OTFS can be implemented as an additional block around an OFDM multicarrier signal~\cite{hadani2018otfsnewgenerationmodulation, cohere2016otfs},  which can dramatically improve link reliability and effective throughput under high mobility scenarios.



Several studies have evaluated OTFS and OFDM modulation techniques in different wireless communication contexts. OTFS modulation is introduced in~\cite{hadani2017otfs}, with theoretical analysis highlighting its robustness to high Doppler conditions compared to OFDM. Subsequent research in~\cite{raviteja2018otfs} focus on significant Doppler shifts and multipath channels, proposing low-complexity iterative detection algorithms for OTFS, also known as Message Passing (MP), presenting a solution for OTFS linear equalizers. Further investigations explore OTFS and OFDM performance using ray-tracing channel models, assessing their comparative Bit Error Rate (BER) and spectral efficiency in sub-6 GHz and mmWave line-of-sight mobility scenarios~\cite{wiffen2018}. This study demonstrated that OTFS can reduce BER significantly if compared to OFDM, emphasizing OTFS's robustness to Doppler effects.

A comprehensive experimental analysis under realistic RF impairments further confirms OTFS's practicality in high-mobility contexts, highlighting considerations for deployment and underscoring performance advantages in realistic scenarios~\cite{Abushattal2022}. Moreover, other assessments presents OTFS's potential as a modulation scheme tailored explicitly for vehicular and high-speed communication systems with significant improvements in reliability and spectral efficiency over existing modulation techniques commonly used in current wireless standards~\cite{Ramachandran2020OTFS}. An overview of OTFS and OFDM is presented in~\cite{Ramachandran2018}, including BER and complexity discussions, emphasizing OTFS's advantages in high-mobility high-speed train scenarios. Further detailed evaluations in vehicle communications demonstrate OTFS's superior BER and enhanced Doppler robustness compared to OFDM~\cite{wiffen2018}.

Additional studies have enriched our understanding of the comparative performance of OTFS and OFDM, particularly in high Doppler and diverse channel conditions. A detailed diversity analysis of OTFS modulation is conducted in~\cite{zhang2019comparison}, revealing its ability to achieve full diversity in both the delay and Doppler domains. The authors implement simulations based on 5G Tapped-Delay-Line (TDL) channel models under both slow and fast fading scenarios. The results indicate that OTFS consistently outperforms OFDM and Single Carrier Frequency Division Multiple Access (SC-FDMA), especially in fast fading environments, underscoring OTFS's effectiveness in exploiting both time and frequency diversity inherent in highly mobile wireless channels.

OTFS and OFDM performance are evaluated in the prospective 6G waveform applications~\cite{mohammadi2022performance}, conducting simulations that compare BER against Signal-to-Noise Ratio (SNR) using 16-QAM modulation. Their findings highlight that OTFS provides superior BER performance in high-speed environments characterized by significant Doppler spreads, reinforcing OTFS as a viable candidate for 6G communication systems where reliability and robustness against Doppler-induced impairments are crucial. Additionally, recent works introduce low-complexity iterative detection methods aimed at reducing OTFS equalization complexity, thereby enhancing practical deployment feasibility~\cite{Thaj_Viterbo2020}.



Despite these contributions, the systematic quantitative analyses with clear boundaries for the requirements and implementations where each modulation scheme is superior are still underexplored. To contribute to the OTFS performance evaluations and to allow a focused investigation on key modulation characteristics, this paper presents a quantitative performance comparison between OTFS and OFDM modulation schemes under the assumption of perfect Channel State Information (CSI). By idealizing the channel estimation process, we isolate and examine the effects of other critical factors on the modulation performance. This approach enables a clear evaluation of both modulations within mobile wireless communication scenarios and at different configurations. The primary contributions of this work are the performance evaluations of the OTFS and OFDM on a diversity of scenarios:


\begin{itemize}
    \item Order modulation schemes evaluation: 4-QAM, 16-QAM, 64-QAM, and 128-QAM.
    \item Mobility scenarios comparison: static ($0$ km/h), mid-mobility ($127$ km/h), and high-mobility ($380$~km/h).
    \item Number of multipath components evaluation: 2, 5, and 8.
    \item Linear and non-linear equalizers for OTFS: Linear Minimum Mean Square Error (LMMSE) and Message Passing method.
\end{itemize}

The rest of the paper is organized as follows. Section~\ref{sec:OFDM_OTFS} presents the fundamental difference between OFDM and OTFS. Section~\ref{sec:system_model} presents the system model used for our evaluations and investigations. In Section~\ref{sec:results}, we present the results and discussion derived from our analyses. Finally, Section~\ref{sec:conclusions} presents the conclusions of our work.

\section{OFDM and OTFS}\label{sec:OFDM_OTFS}


This section presents the basic theory of OFDM and OTFS modulations and their key mathematical concepts, such as Fourier-based transformations and modulation parameters.

\subsection{Orthogonal Frequency Division Multiplexing (OFDM)}

OFDM is the dominant modulation scheme extensively used in contemporary wireless communication standards such as LTE, Wi-Fi, and 5G NR due to its inherent simplicity, robustness against multipath fading, and efficient spectrum utilization. OFDM operates basically by dividing the available spectrum into multiple orthogonal subcarriers, which are modulated independently using an Inverse Fast Fourier Transform (IFFT)~\cite{rappaport2020wireless}. 





The information symbols are defined in a time-frequency domain matrix, $\boldsymbol{X}_{TF_{ofdm}}[m,n]$. These symbols are passed to the time domain and vectorized through the Heisenberg transform, resulting in the transmitted signal described by~\cite{hong2022delay}~as

\begin{equation}
    \label{eq_OFDM}
    \resizebox{0.9\columnwidth}{!}{$
    s_{_{ofdm}}(t) = \displaystyle\sum_{n=0}^{N-1} \sum_{m=0}^{M-1} X_{TF_{ofdm}}[m,n] g(t-nT) e^{j2 \pi m \Delta f(t-nT)} \text{,}
    $}
\end{equation}

\noindent where $M$, $N$, $\Delta f$, $T$ and $g(t)$ represent the number of subcarriers, the number of OFDM symbols, the sampling frequency, the OFDM symbol duration, and the pulse shaping waveform for the continuous-time signal, respectively. Despite its widespread use, OFDM is highly susceptible to inter-carrier interference under high Doppler shifts, limiting its applicability in high-speed mobility scenarios such as vehicular or aerial communications.


The received signal on the $m$-th subcarrier can be modeled as~\cite{hong2022delay}

\begin{equation}\label{eq:vector_form}
    \boldsymbol{y_m} = \boldsymbol{ H_m x_m } + \boldsymbol{ \omega_m },
\end{equation}


\noindent where $\boldsymbol{y_m}$ is the received symbol vector, $\boldsymbol{H_m}$ is the channel frequency response matrix on subcarrier $m$, $\boldsymbol{x_m}$ is a version of the transmitted symbol vector, $s_{_{ofdm}}(t)$, but serialized directly from the time-frequency domain, and $\boldsymbol{\omega_m}$ is the Additive White Gaussian Noise (AWGN) vector with variance $\sigma^2$. The LMMSE equalizer estimates the transmitted symbol as~\cite{proakis2008digital}
\begin{equation}
    \boldsymbol{ \hat{x}_m^{\text{LMMSE}} } = \left(| \boldsymbol{H_m} |^2 + \sigma^2 \boldsymbol{I}\right) \boldsymbol{H_m^* y_m} \text{,} 
\end{equation}

\noindent where $\boldsymbol{I}$ is the identity matrix. This formulation demonstrates how LMMSE balances noise enhancement and Inter-Symbol Interference (ISI) mitigation.




\subsection{Orthogonal Time Frequency Space (OTFS)}

OTFS modulation has been recently proposed as a robust solution to address the challenges posed by high mobility scenarios. Unlike OFDM, OTFS transforms the channel representation from the time-frequency domain into the delay-Doppler domain, significantly simplifying the channel's complexity by converting time-varying multipath channels into nearly static representations. This facilitates enhanced robustness and reliability in rapidly varying channels.


\begin{figure*}[!t]
  \centering  
  \includegraphics[width=\textwidth]{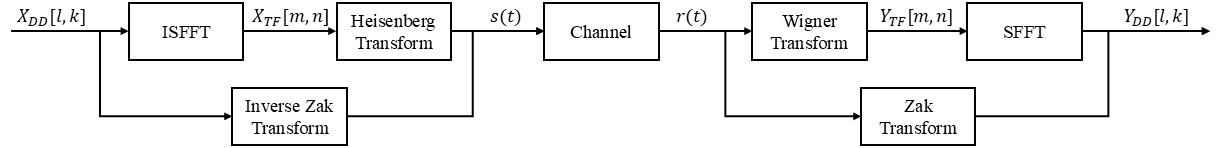}
  \caption{OTFS modulation scheme.}
  \label{fig:otfs_diagram}
\end{figure*}





OTFS modulation can be implemented by two basic schemes, as presented in Fig.~\ref{fig:otfs_diagram}. In the first case, the information symbols $\boldsymbol{X}_{DD}[l,k]$, defined in the delay-Doppler domain, are mapped onto the time-frequency plane through an Inverse Symplectic Finite Fourier Transform (ISFFT), represented mathematically as~\cite{hong2022delay}

\begin{equation}
    \boldsymbol{X}_{\text{TF}}[m,n] = \frac{1}{\sqrt{NM}} \sum_{k=0}^{N-1} \sum_{l=0}^{M-1} \boldsymbol{X}_{DD}[l,k] e^{j 2 \pi (\frac{nk}{N} - \frac{ml}{M})} \text{,}
\end{equation}

\noindent for $l = 0, \cdots, M-1$ and $k = 0, \cdots, N-1$. The $M$ and $N$ represent the number of subcarriers and OTFS symbols, respectively. The time-frequency domain representation $\boldsymbol{X}_{TF}[m,n]$ is subsequently converted into a time-domain signal, $s(t)$, using the Heisenberg transform~\cite{hadani2017otfs}, as in OFDM modulation. At the receiver, the reverse transformation (Wigner transform followed by a SFFT) converts the received signal back into the delay-Doppler domain.

Alternatively, in order to reduce computational cost, OTFS modulation can be performed through the Inverse Discrete Zak Transform (IDZT) that maps the information symbols in the delay-Doppler domain onto the discrete time domain ~\cite{hong2022delay}, that is

\begin{equation}
    s[q] = s[l+nM] = \frac{1}{\sqrt{N}} \sum_{k=0}^{N-1} X_{DD}[l,k] e^{j\frac{2 \pi}{N} nk} \text{.}
\end{equation}

After that, a Digital-to-Analog (DA) converter is used to form the transmitted signal in the continuous time domain, $s(t)$. Similarly to  \eqref{eq:vector_form}, the received signal $\boldsymbol{r}$ can be modeled as

\begin{equation}\label{eq:matrix_form}
    \boldsymbol{r} = \boldsymbol{G s} + \boldsymbol{\omega} \text{,}
\end{equation}

\noindent where $\boldsymbol{G}$ in $\mathbb{C}^{NM \times NM}$ is the channel matrix in the delay-time domain, $\boldsymbol{s}$ in $\mathbb{C}^{NM \times 1}$ is the transmitted symbol vector, and $\boldsymbol{\omega}$ is the complex Gaussian noise. The received signal is then passed into the discrete delay-Doppler domain through an Analog-to-Digital (AD) converter, followed by the Discrete Zak Transform (DZT).


In order to estimate the received symbols, the LMMSE equalizer aims to minimize the mean squared error between the transmitted and estimated symbols. The LMMSE estimate of the transmitted symbols vector is given by~\cite{proakis2008digital}

\begin{equation}
    \boldsymbol{\hat{x}}^{LMMSE} = \left(\boldsymbol{H}^* \boldsymbol{H} + \sigma^2 \boldsymbol{I}\right)^{-1} \boldsymbol{H}^* \boldsymbol{y} \text{,}
\end{equation}

\noindent where $\boldsymbol{y}$ and $\boldsymbol{H}$ are the received signal vector and the channel matrix in the delay-Doppler domain, respectively.

This solution mitigates ISI and limits noise amplification by regularizing the inversion. Due to the typically large dimensions and sparsity of $\boldsymbol{H}$, in practical OTFS scenarios, efficient implementations often rely on iterative and non-linear solutions such as MP detectors to avoid direct matrix inversion.

The MP equalizer iteratively updates the posterior probabilities of transmitted symbols. Specifically, MP computes messages between symbols using factor graphs defined by the channel model. Based on Equation~\eqref{eq:matrix_form}, the Maximum a Posteriori Probability (MAP) detection rule for estimating the transmitted signals is given by~\cite{raviteja2018otfs}

\begin{equation}
    \boldsymbol{\hat{x}} = \underset{\boldsymbol{x} \in \mathbb{C}^{NM \times 1}}{\arg\max} \text{Pr} (\boldsymbol{x} | \boldsymbol{y}, \boldsymbol{H}) \text{.}
\end{equation}

Thus, the algorithm updates iteratively the symbols' beliefs using local probabilistic message exchanges from the channel~$\boldsymbol{H}$. Compared to the LMMSE, MP offers improved detection in highly dispersive channels by leveraging the sparse delay-Doppler channel matrix. However, this performance gain comes at the cost of higher computational complexity due to the iterative process. Moreover, the convergence of MP is not always guaranteed and might depend on the channel sparsity and the number of iterations~\cite{raviteja2018otfs}.



\section{System Model}\label{sec:system_model}

\begin{figure}[H]
  \centering
  \includegraphics[width=\columnwidth]{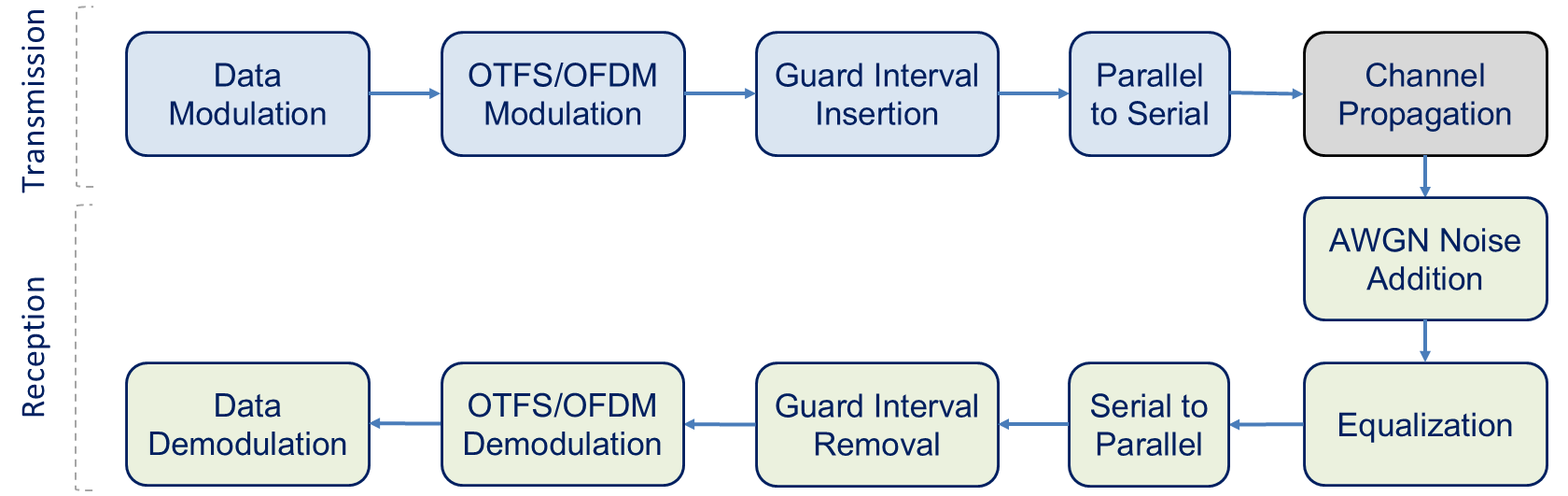}
  \caption{System model overview.}
  \label{fig:Overview}
\end{figure}

This section describes the methodology and system model employed to quantitatively compare the performance of OTFS and OFDM modulation schemes in mobile wireless communication scenarios, as summarized in Fig.~\ref{fig:Overview}. Our evaluation approach includes defining the simulation configuration, the evaluated scenarios, the communication system parameters, and the metric used for performance analysis.

The simulation considers perfect CSI at the receiver, zero padding is used for guard intervals, and the number of subcarriers and OTFS symbols are equal to $32$, as presented in Table~\ref{tab:OTFS_params}. The wireless channel is generated synthetically in the delay-Doppler domain using a Tapped Delay Line model. Each channel realization is composed of a discrete number of paths (\textit{P}), each with randomly assigned normalized delay ($l_{i}$) and Doppler shift ($k_{i}$) indices, and sampled random complex path gains ($h_{i}$). Unless otherwise stated in the Section~\ref{sec:results}, the parameters remain as defined in Table~\ref{tab:OTFS_params}.

\begin{table}[htbp]
\centering
\caption{Simulation Parameters.}
\begin{tabular}{lc}
\hline
\textbf{Parameter}                             & \textbf{Value}          \\ \hline
Number of Subcarriers                      & $M=32$                      \\ 
Number of OTFS Symbols                      & $N=32$                      \\ 
Subcarrier Spacing                 & $\Delta f=15~kHz$                  \\ 
Sampling Frequency / Signal Bandwidth                & $f_s=B=480~kHz$                  \\ 
OTFS Symbol Duration                      & $T=66.67\,\mu s$           \\ 
Frame Duration                          & $T_f=2.13~ms$               \\ 
Doppler Resolution                 & $\Delta v=468.75~Hz$               \\ 
Delay Resolution                        & $T_s=2.08\,\mu s$               \\ 
Maximum Doppler Shift      & $f_{d,\text{max}}=1.41~kHz$                 \\ 
Maximum Delay Spread      & $\tau_{\text{max}}=8.33\,\mu$s           \\ 
Maximum Normalized Doppler Shift      & $k_{\text{max}}=3$                 \\ 
Maximum Normalized Delay Spread      & $l_{\text{max}}=4$           \\ 
Maximum Velocity      & $v_{\text{max}}=380~km/h$           \\ 
Number of Multipath Components      & $P=5$           \\ 
Channel Coefficients       &  $h_{i}\sim \textit{CN}(0, 1/P)$           \\ 
Modulation Order       &  $4-QAM$           \\ 
Equalizer Type       &  $LMMSE$           \\ 
Carrier Frequency                       & $f_c=4~GHz$                   \\ Quantity of Monte Carlo Trials                      & $10^4$                   \\ \hline

\end{tabular}
\label{tab:OTFS_params}
\end{table}



OTFS and OFDM systems utilize identical underlying parameters to ensure a fair comparison. The simulations operate at a carrier frequency of $4$ GHz, common in contemporary mobile networks. We maintain a consistent bandwidth and subcarrier spacing of $15$ kHz for both modulation schemes. For the OTFS system, we adopt a delay-Doppler grid configuration comprising $M$ subcarriers and $N$ OTFS symbols and the Zak transform scheme. We assess modulation orders, including 4-QAM, 16-QAM, 64-QAM, and 128-QAM, to explore how higher modulation schemes' performance are affected by high Doppler conditions.

\subsection{Evaluation Scenarios}

We systematically compare the performance of OFDM and OTFS over different scenarios, characterized by multiple mobility levels, by varying the maximum user velocity while keeping the same other physical channel model characteristics. The scenarios are static, intermediate mobility, and high mobility. Since we are using a non-fractional delay-Doppler grid, the Doppler indices are integers, and when multiplied by the Doppler resolution, we obtain the Doppler values that are correlated to velocity values. Specifically, the static scenario corresponds to a speed of $0$~km/h, where the channel can be considered almost time-invariant over the frame duration ($2.13$~ms) and Doppler effects are negligible. The intermediate mobility scenario corresponds to a maximum speed of $127$~km/h, leading to a moderate Doppler spread introducing selectivity and ICI in the OFDM system. Finally, the high mobility scenario models a maximum speed of $380$~km/h, resulting in severe Doppler effects and fast channel variations within a single frame.

\subsection{Performance Metrics}

We quantitatively evaluate both modulation schemes using bit error rate as our performance metric. BER is defined as the ratio of erroneous bits received over the total number of transmitted bits. BER simulations are performed across a range of SNR values to assess robustness against noise and Doppler-induced interference. Additionally, we explore the boundaries by identifying potential crossover points and mobility thresholds at which one modulation scheme might outperform the other significantly. The simulation results are averaged over $10^4$ Monte Carlo trials to ensure statistical reliability and accuracy of the performance evaluation.

\section{Results}\label{sec:results} 

In our analysis, we consider four dimensions: modulation order, mobility scenarios, number of multipath components, and equalization technique.

\subsection{Modulation Order}

\begin{figure}[H]
  \centering
  \includegraphics[width=\columnwidth, trim=1cm 0.2cm 1.5cm 0.7cm, clip]{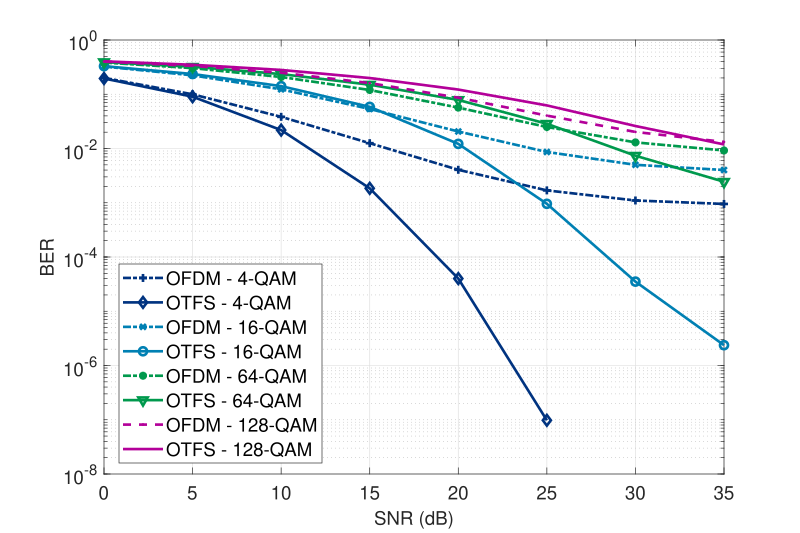}
  \caption{BER performance comparison for multiple modulation order (LMMSE, P = 5 and 380~km/h).}
  \label{fig:modulation_order}
\end{figure}

We evaluate the impact of different modulation orders, specifically $4, 16, 64, 128$, on the BER performance of OTFS and OFDM. Fig.~\ref{fig:modulation_order} presents OTFS outperforming OFDM for modulation orders of $4$ and $16$, while presenting crossover performance points and trade-offs across orders $64$ and $128$. The performance gap increases between them with higher SNR values and lower modulation order. For instance, at an order of $4$, OTFS has significantly lower BER compared to OFDM. While for an order of $64$, OTFS outperforms OFDM only for SNR values higher than $26$ dB, approximately.

\subsection{Mobility Scenarios} 

\begin{figure}[H]
  \centering
  \includegraphics[width=\columnwidth, trim=1cm 0.2cm 1.5cm 0.7cm, clip]{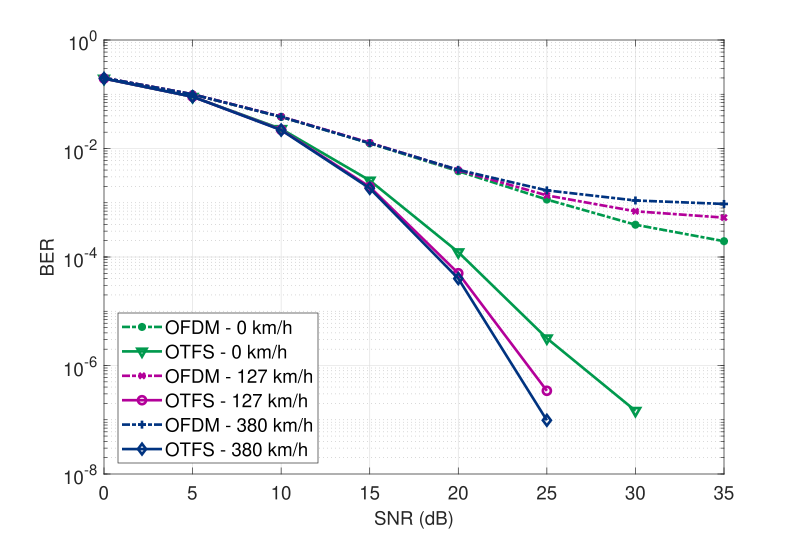}
  \caption{BER performance comparison for multiple mobility scenarios (LMMSE, P = 5 and 4-QAM).}
  \label{fig:mobility_scenarios}
\end{figure}

Fig.~\ref{fig:mobility_scenarios} presents the mobility scenarios performance comparisons. The performance gap clearly increases with the mobility levels, indicating superior robustness of OTFS against high mobility scenarios. At high speeds, OFDM degrades in performance due to severe inter-carrier interference from Doppler spread, while OTFS increases performance due to its delay-Doppler domain nature, which explores the channel diversity. At high speed ($380$ km/h), OTFS reaches BER reductions of almost an order of magnitude compared to OFDM at $15$ dB SNR, and approximately four orders of magnitude at $25$ dB SNR. Even in the static channel, OTFS has superior performance compared to OFDM because it spreads the information symbol across the entire delay-Doppler grid, increasing robustness against channel frequency selectivity. It is worth noting that OFDM variants, such as SC-FDMA, have equivalent performance to OTFS in this type of channel~\cite{zhang2019comparison}.


\subsection{Number of Multipaths Components}

\begin{figure}[H]
  \centering
  \includegraphics[width=\columnwidth, trim=1cm 0.2cm 1.5cm 0.7cm, clip]{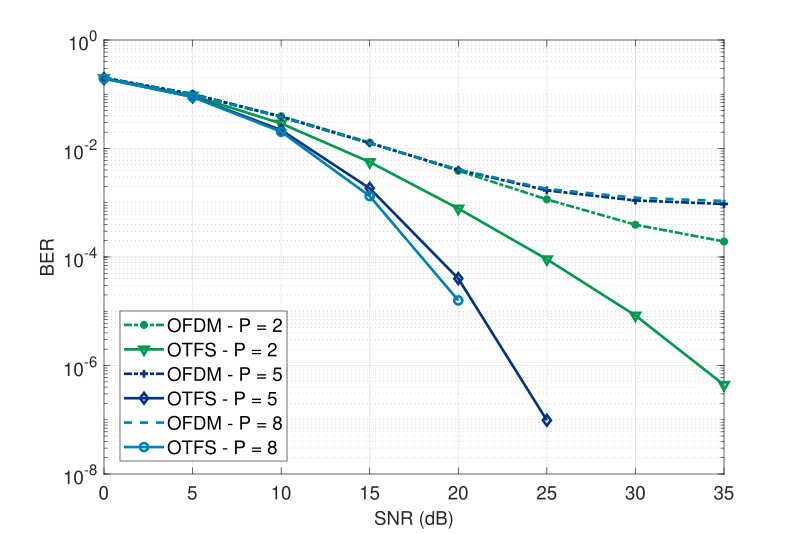}
  \caption{BER performance comparison for different numbers of paths (LMMSE, 4-QAM and 380~km/h).}
  \label{fig:number_of_paths}
\end{figure}

We assess the effect of multipath variation by changing the number of channel paths, $P = 2,5,8$. Fig.~\ref{fig:number_of_paths} presents robust performance for OTFS. Higher the number of multipath components, higher is the performance gap between OTFS and OFDM. At $P=2$, for $25$ dB SNR, OTFS presents the BER more than one order of magnitude lower than OFDM. While for $P=8$, the difference is approximately four orders of magnitude.

\subsection{Equalizers}

\begin{figure}[H]
  \centering
  \includegraphics[width=\columnwidth, trim=1cm 0.2cm 1.5cm 0.7cm, clip]{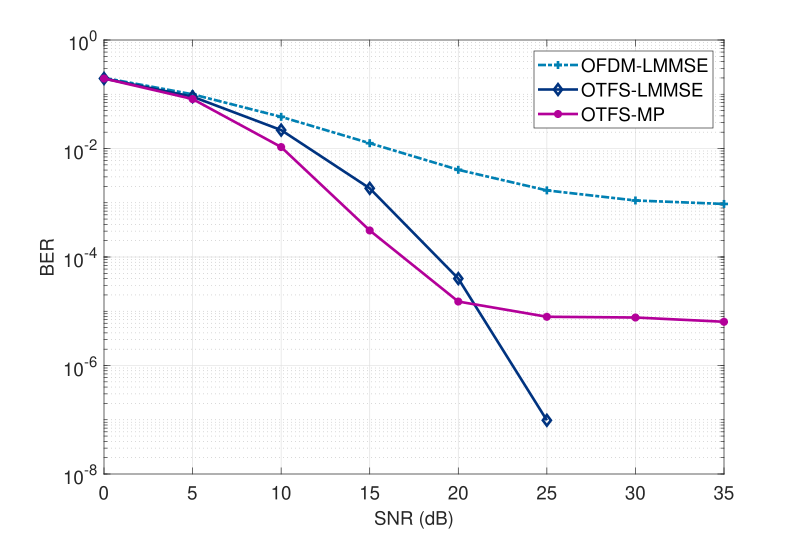}
  \caption{BER performance comparison for different equalizers (P = 5, 4-QAM and 380~km/h).} 
  \label{fig:equalization}
\end{figure}

We compare OTFS and OFDM using LMMSE equalization against OTFS with MP equalization. Fig.~\ref{fig:equalization} shows that OTFS-MP outperforms OTFS-LMMSE for low and mid-range SNR values (lower than $21$ dB). Although LMMSE is computationally simpler than MP equalization, the MP equalizer improves symbol detection by leveraging the channel sparsity and iterative method, justifying its additional complexity in low SNR scenarios.

\section{Conclusions}\label{sec:conclusions}

This paper presents a performance comparison between OTFS and OFDM modulation schemes over a variety of wireless communication scenarios. Through simulations, we examined their performance across four relevant dimensions: modulation order, mobility levels, number of multipath components, and equalization technique. The study explores all four aspects in a single unified framework. These evaluations allow us to contribute to a clear comprehension of where each modulation scheme and its configurations excel. 

Our results show that OTFS outperforms OFDM in scenarios characterized by high mobility and multipath propagation, mainly at low and mid modulation orders ($4, 16$). OTFS demonstrates strong robustness to Doppler spread by leveraging the delay-Doppler domain, where the channel becomes approximately invariant. This allows OTFS to remain low BER even in extreme vehicular and high-speed train environments. However, the results show that such resiliency is less significant at high modulation orders ($64, 128$), where OTFS only outperforms OFDM across high SNR scenarios. Moreover, the evaluation of equalizers reveals that the MP scheme presents a gain over OTFS using LMMSE detections, particularly at low/mid SNR range values (between $0-21$ dB).

Future work includes the implementation of OFDM variants and other equalization techniques. Furthermore, Peak-to-Average Power Ratio (PAPR) and computational cost evaluations can be explored.



\bibliographystyle{IEEEtran}
\bibliography{references}


\end{document}